\documentclass[a4paper,11pt]{article}
\pdfoutput=1 % if your are submitting a pdflatex (i.e. if you have
\usepackage{jcappub} % for details on the use of the package, please
\usepackage{xcolor}
\usepackage[T1]{fontenc} % if needed

\usepackage{bm}
\usepackage{epsfig}
\usepackage[lofdepth,lotdepth,caption=false]{subfig}

\usepackage{graphicx}% Include figure files

\usepackage{float}
\usepackage{graphicx}
\usepackage{amsmath,amssymb}
\usepackage{textcomp}
\usepackage{gensymb}
\usepackage[utf8]{inputenc}
\usepackage[T1]{fontenc}
\usepackage{setspace}
\usepackage{hyperref}

\def\beq{\begin{equation}}
\def\enq{\end{equation}}
\def\bea{\begin{eqnarray}}
\def\ena{\end{eqnarray}}

\begin{document}

\title{Constraining high-energy neutrinos from choked-jet supernovae with IceCube high-energy starting events}

\author[a]{Arman Esmaili,}

\author[b,c]{Kohta Murase}

\affiliation[a]{Departamento de F\'isica, Pontif\'icia Universidade Cat\'olica do Rio de Janeiro, Rio de Janeiro 22452-970, Brazil}
\affiliation[b]{Department of Physics; Department of Astronomy \& Astrophysics; Center for Particle and Gravitational Astrophysics, The Pennsylvania State University, University Park, Pennsylvania 16802, USA}
\affiliation[c]{Yukawa Institute for Theoretical Physics, Kyoto, Kyoto 606-8502, Japan}

\abstract{
Different types of core-collapse supernovae (SNe) have been considered as candidate sources of high-energy cosmic neutrinos. Stripped-envelope SNe, including energetic events like hypernovae and super-luminous SNe, are of particular interest. They may harbor relativistic jets, which are capable of explaining the diversity among gamma-ray bursts (GRBs), low-luminosity GRBs, ultra-long GRBs, and broadline Type Ib/c SNe. Using the six-year IceCube data on high-energy starting events (HESEs), we perform an unbinned maximum likelihood analysis to search for spatial and temporal coincidences with 222 samples of SNe Ib/c. We find that the present data are consistent with the background only hypothesis, by which we place new upper constraints on the isotropic-equivalent energy of cosmic rays, ${\mathcal E}_{\rm cr}\lesssim{10}^{52}~{\rm erg}$, in the limit that all SNe are accompanied by on-axis jets. Our results demonstrate that not only upgoing muon neutrinos but also HESE data enable us to constrain the potential contribution of these SNe to the diffuse neutrino flux observed in IceCube. We also discuss implications for the next-generation neutrino detectors such as IceCube-Gen2 and KM3Net.}

\maketitle
\date{\today}

%%%%%%%%%%%%%%%%%%%%%%%%%%%%%%%%%%%%%%%%%%%%%%%%%%
%%%%%%%%%%%%%%%%%%%%%%%%%%%%%%%%%%%%%%%%%%%%%%%%%%

\section{Introduction}
\doublespacing
The recent discoveries of high-energy cosmic neutrinos~\cite{Aartsen:2013bka,Aartsen:2013jdh} and gravitational waves~\cite{Abbott:2016blz,TheLIGOScientific:2017qsa} have opened the new era of multi-messenger astroparticle physics. 
Powerful explosive phenomena such as gamma-ray bursts (GRBs) and supernovae (SNe), including low-luminosity GRBs and transrelativistic SNe, are among the candidate sources of IceCube neutrinos~\cite{Waxman:1997ti,Meszaros:2001ms,Murase:2006mm,Gupta:2006jm,Murase:2009pg}, and expected to have bright electromagnetic counterparts at different wavelengths. In particular, gamma-ray emission of long-duration GRBs is thought to originate from relativistic jets launched at the death of massive stars. The jet has to successfully break out from is progenitor star for the gamma-ray emission to be observed. However, it is also natural for the jet to fail to penetrate the progenitor star, if the jet power is not high enough or the star has an extended structure~\cite{Matzner:2002ti,Suwa:2010ze}. Such failed GRBs could be seen as low-luminosity (LL) GRBs or energetic SNe Ib/c with a relativistic component of the SN ejecta~\cite{Campana:2006qe,Soderberg:2006vh,Margutti:2014gha}. Such choked-jet SNe are likely to possess a key link between GRBs and SNe, and the relativistic jet may play an important role in making the diversity among Type Ib/c SNe, including broadline Type Ib/c SNe and super-luminous Ic SNe~\cite{Thompson:2004wi,Chakraborti:2014dha,Nakar:2015tma,Barnes:2017hrw}. 

Non-thermal properties of observed GRBs suggest that particles can be accelerated by a jet up to high energies. However, the electromagnetic emission cannot be directly observed if it is choked. High-energy neutrino observations provide a unique opportunity to probe the choked jet hidden inside a star~\cite{Meszaros:2001ms,Razzaque:2004yv,Ando:2005xi,Iocco:2007td}, and enable us to study particle acceleration mechanisms in a photon-rich environment~\cite{Murase:2013ffa}. In addition, not only the jet emission but also subsequent shock breakout emission may lead to high-energy neutrino and gamma-ray signatures in the presence of a dense circumstellar material~\cite{Murase:2010cu,Katz:2011zx,Kashiyama:2012zn,Murase:2017pfe}.  

The main origin of high-energy cosmic neutrinos observed in IceCube is unknown (see a review~\cite{Halzen:2016gng}). LL GRBs and choked-jet SNe are suggested as the main candidate sources of IceCube neutrinos~\cite{Murase:2006mm,Murase:2013ffa}, and their contribution to the diffuse neutrino flux have been extensively studied~\cite{Liu:2011cua,Bhattacharya:2014sta,Senno:2015tsn,Tamborra:2015fzv,Denton:2017jwk,Denton:2018tdj,He:2018lwb,Boncioli:2018lrv}. They are among the classes of gamma-ray dark or hidden neutrino sources, and seem necessary to explain the diffuse neutrino flux in medium-energy (10-100~TeV) range~\cite{Murase:2015xka}. Contrary to canonical GRBs whose contribution to the background flux is strongly constrained~\cite{Abbasi:2012zw,Aartsen:2014aqy,Adrian-Martinez:2016xij,Aartsen:2017wea}, these classes of SNe and GRBs are not accompanied by bright gamma-ray emission, so the existing stacking analyses do not give any strong constraint. The absence of clustering in the arrival distribution of neutrinos gives a limit on the rate density, $\gtrsim10-100~{\rm Gpc}^{-3}~{\rm yr}^{-1}$~\cite{Senno:2016bso,Aartsen:2018fpd}, but LL GRBs and SNe with long durations of $\gtrsim100-1000$~s are allowed by this constraint. More dedicated searches focusing on these SNe are necessary. Ref.~\cite{Senno:2017vtd} recently provided a new constraint on high-energy emission from Type Ib/c SNe, by searching for spatial and temporal correlations between through-going muon neutrinos and these SNe. In this work, for the first time, we present results of a stacking analysis for SNe, using six-year high-energy starting events (HESEs) observed in IceCube. Although the angular resolution of HESE events is not as good as that of the muon track events, thanks to the temporal information, we can still obtain a useful limit on the energy of cosmic rays produced in SNe. This limit can be translated into a constraint on the SN contribution to the diffuse neutrino flux, which is useful for us to address the question about the origin of high-energy cosmic neutrinos. 

The paper is organized as follows. 
In Sec.~\ref{data}, we describe the neutrino and SN data used in our likelihood analysis as well as the details of the method. 
Sec.~\ref{limit}, based on the results of the likelihood analysis, shows the constraints on the amount of cosmic-ray energy.  
In Sec.~\ref{discussion}, we compare our results to other limits obtained in the literature. 
In Sec.~\ref{summary}, we summarize our findings and discuss prospects for future observations. 

%%%%%%%%%%%%%%%%%%%%%
%%%%%%%%%%%%%%%%%%%%%
\section{Data Selection and Analysis Method}\label{data}
%%%%%%%%%%%%%%%%%%%%%
%%%%%%%%%%%%%%%%%%%%%

%%%%%%%%%%%%%%%%%%%%%%%%%%%%%%
%%%%%%%%%%%%%%%%%%%%%%%%%%%%%%
\subsection{Neutrino and SN data\label{sec:data}}
%%%%%%%%%%%%%%%%%%%%%%%%%%%%%%
%%%%%%%%%%%%%%%%%%%%%%%%%%%%%%

%%%%%%%%%            figure 1           %%%%%%%%%%%%%
%%%%%%%%%%%%%%%%%%%%%%%%%%%%%%%%%
\begin{figure}[h!]
\centering
\subfloat[]{
\includegraphics[width=0.5\textwidth]{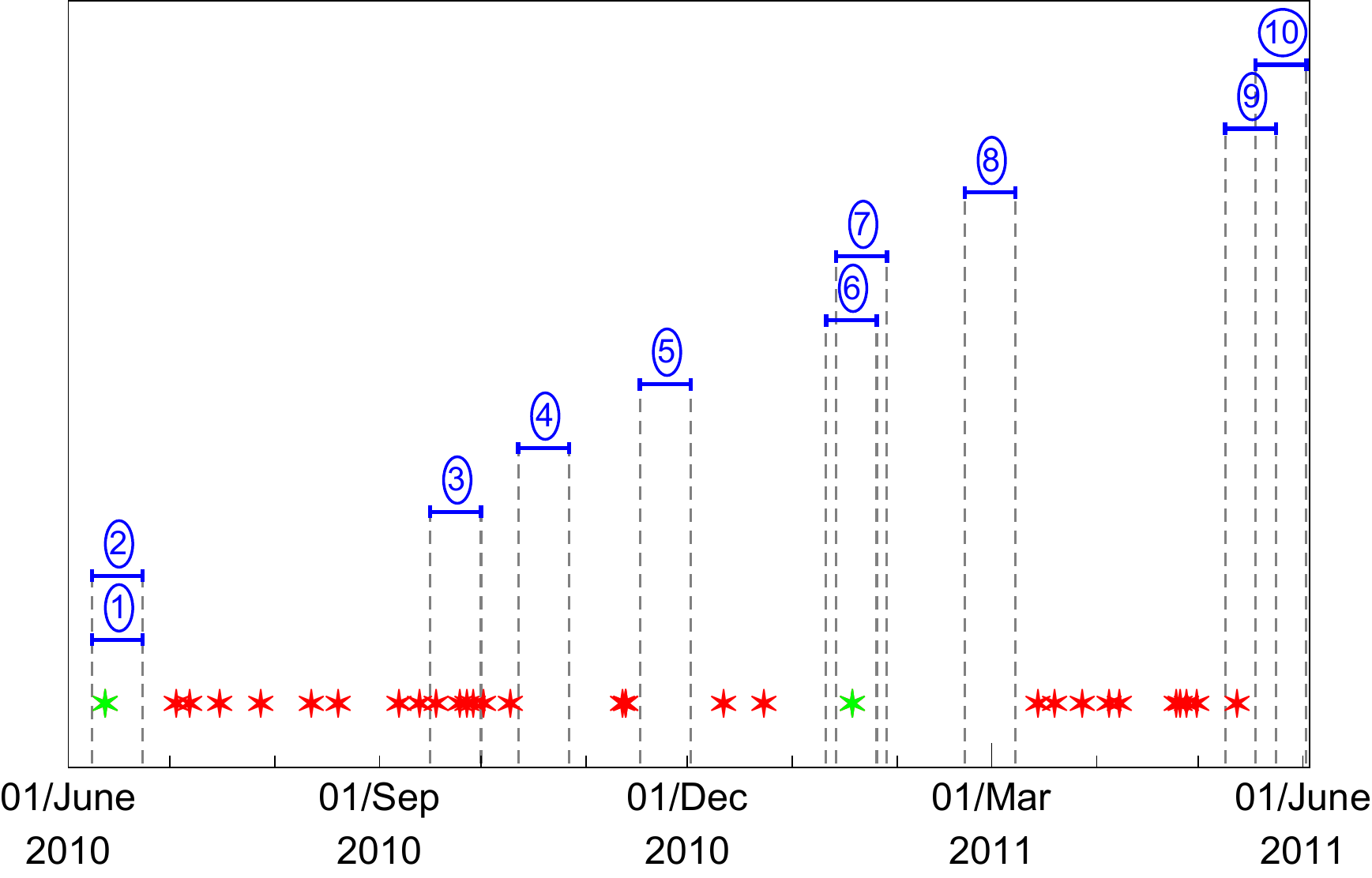}
\label{fig:dndx}
}
\subfloat[]{
\includegraphics[width=0.5\textwidth]{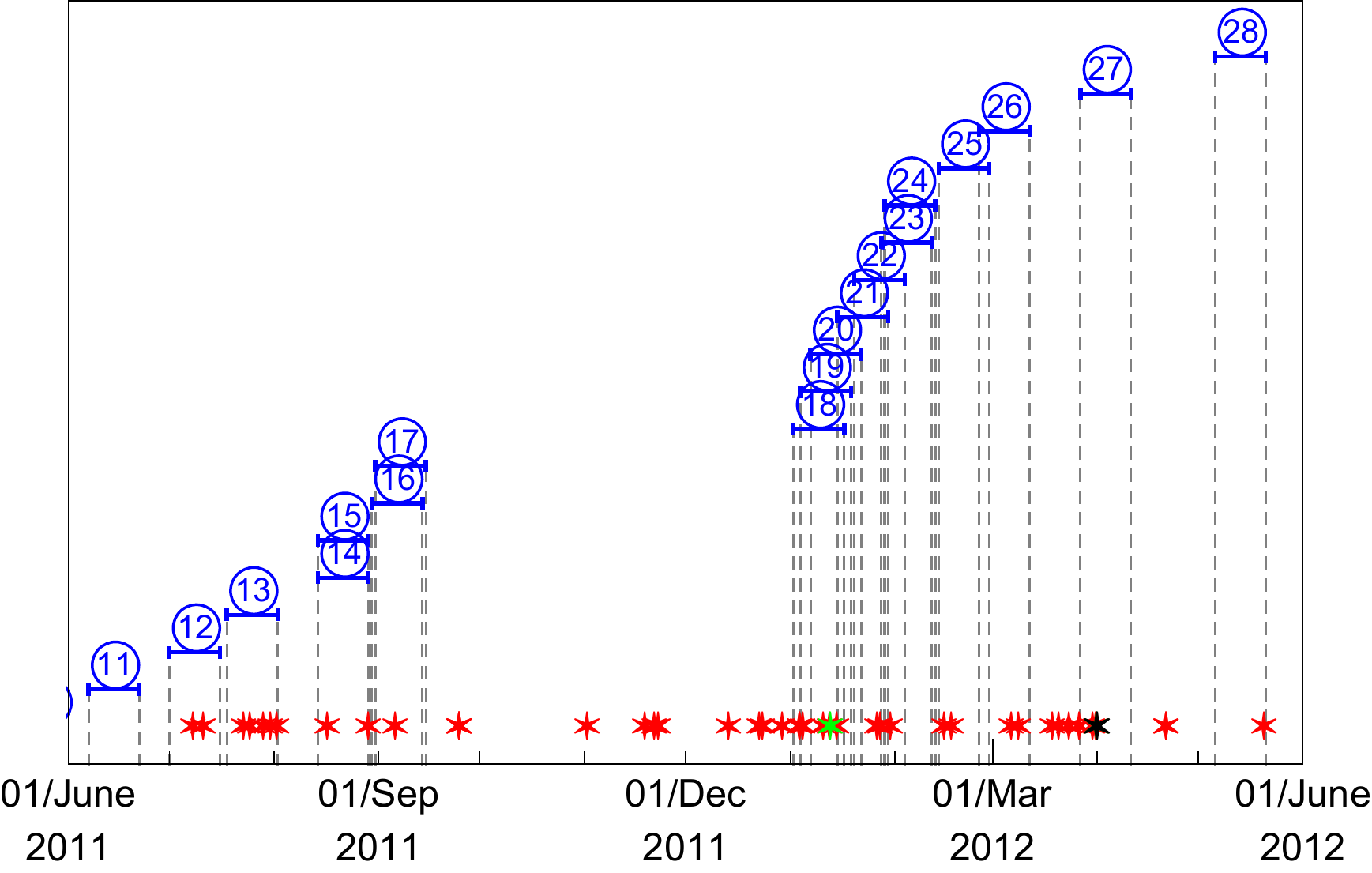}
\label{fig:lum}
}
\quad
\subfloat[]{
\includegraphics[width=0.5\textwidth]{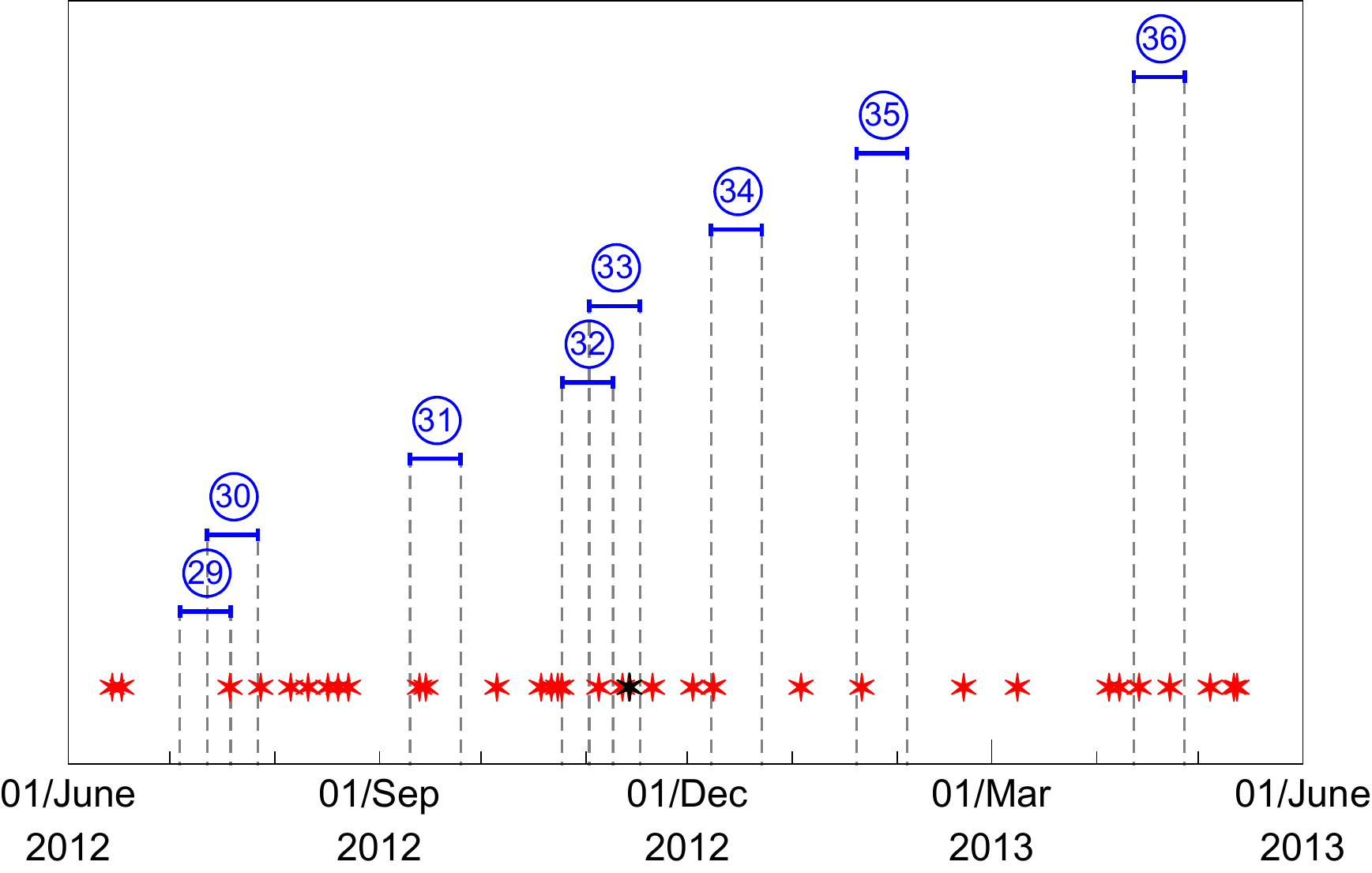}
\label{fig:dndx}
}
\subfloat[]{
\includegraphics[width=0.5\textwidth]{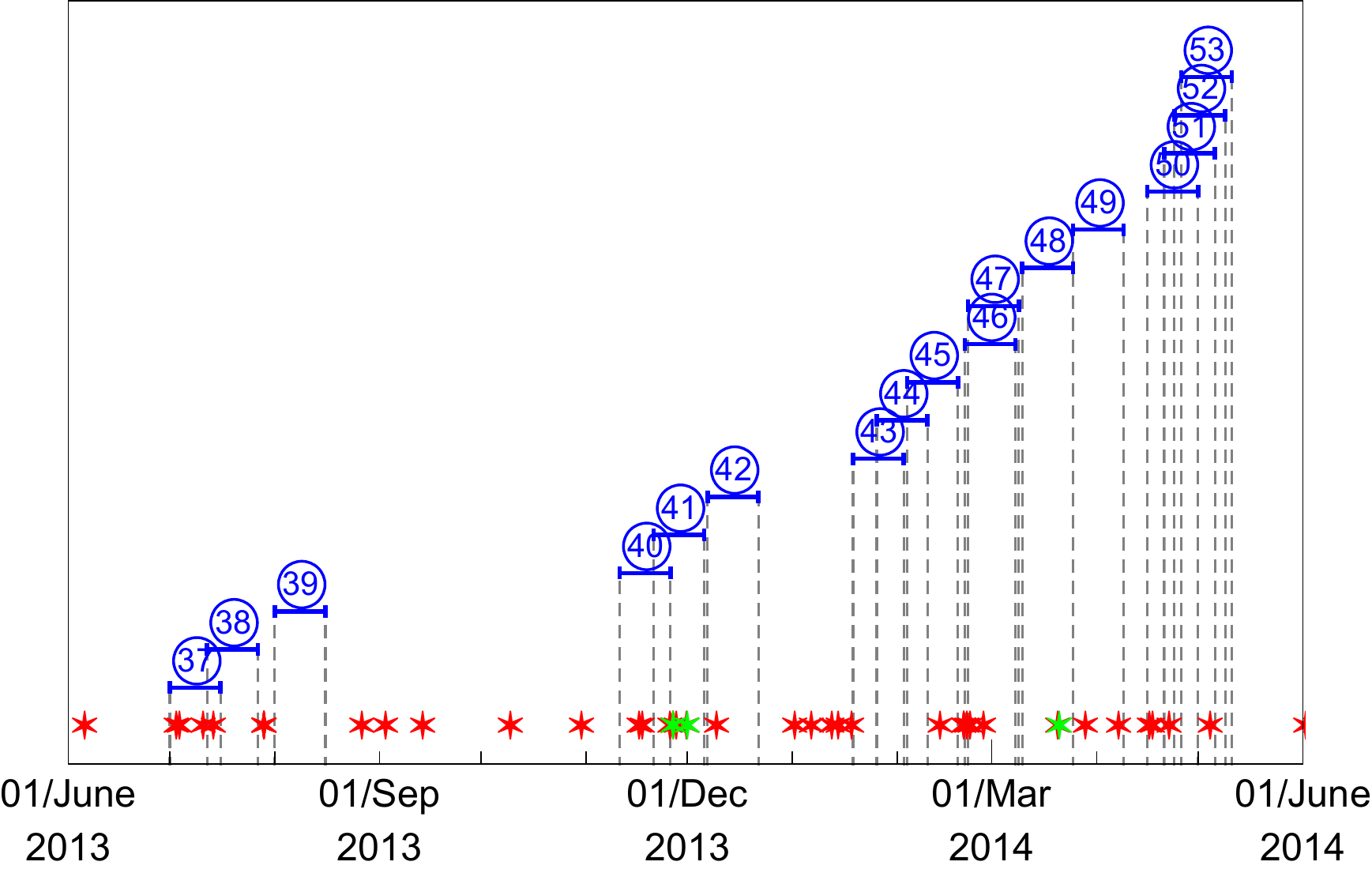}
\label{fig:lum}
}
\quad
\subfloat[]{
\includegraphics[width=0.5\textwidth]{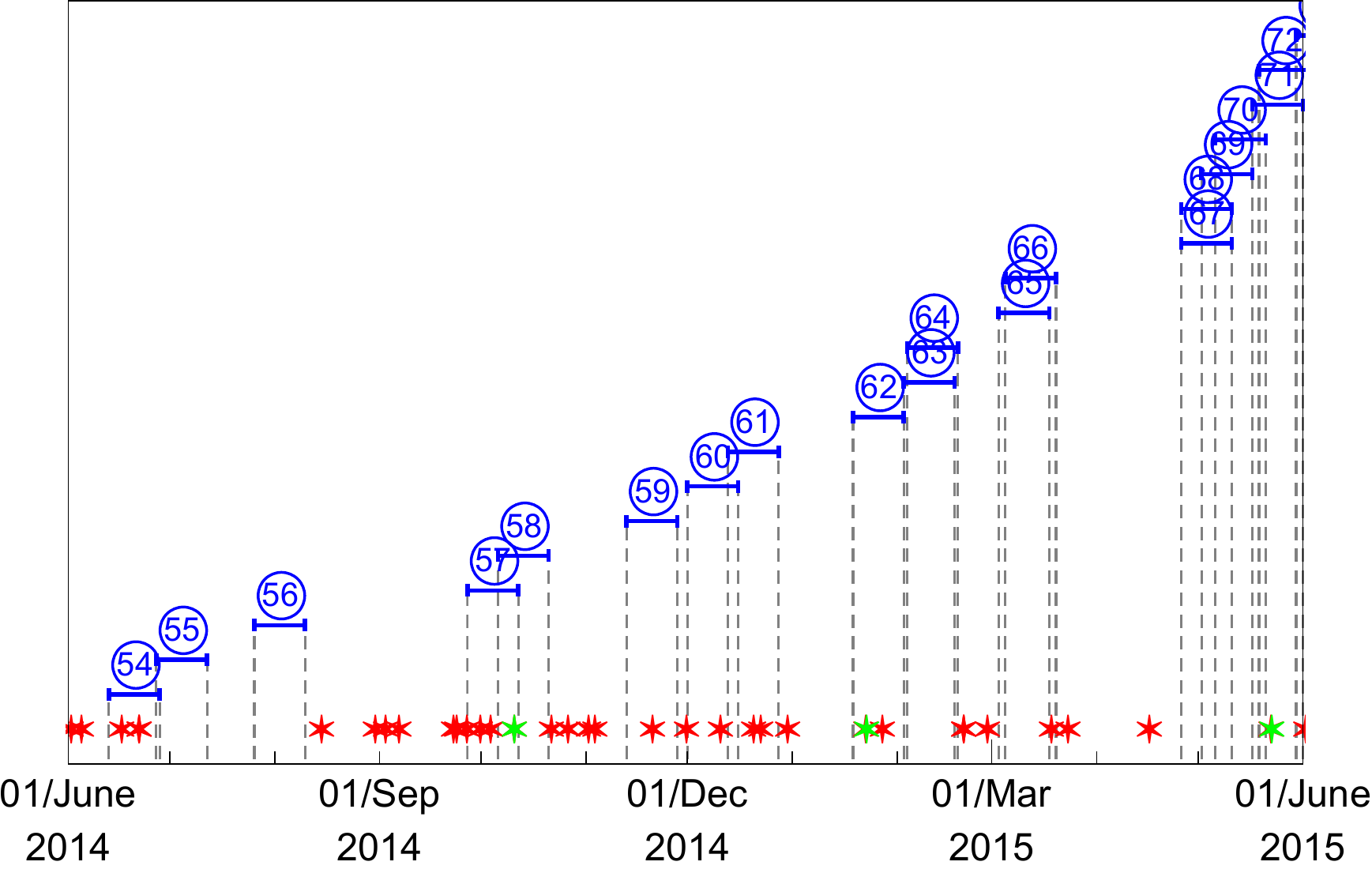}
\label{fig:dndx}
}
\subfloat[]{
\includegraphics[width=0.5\textwidth]{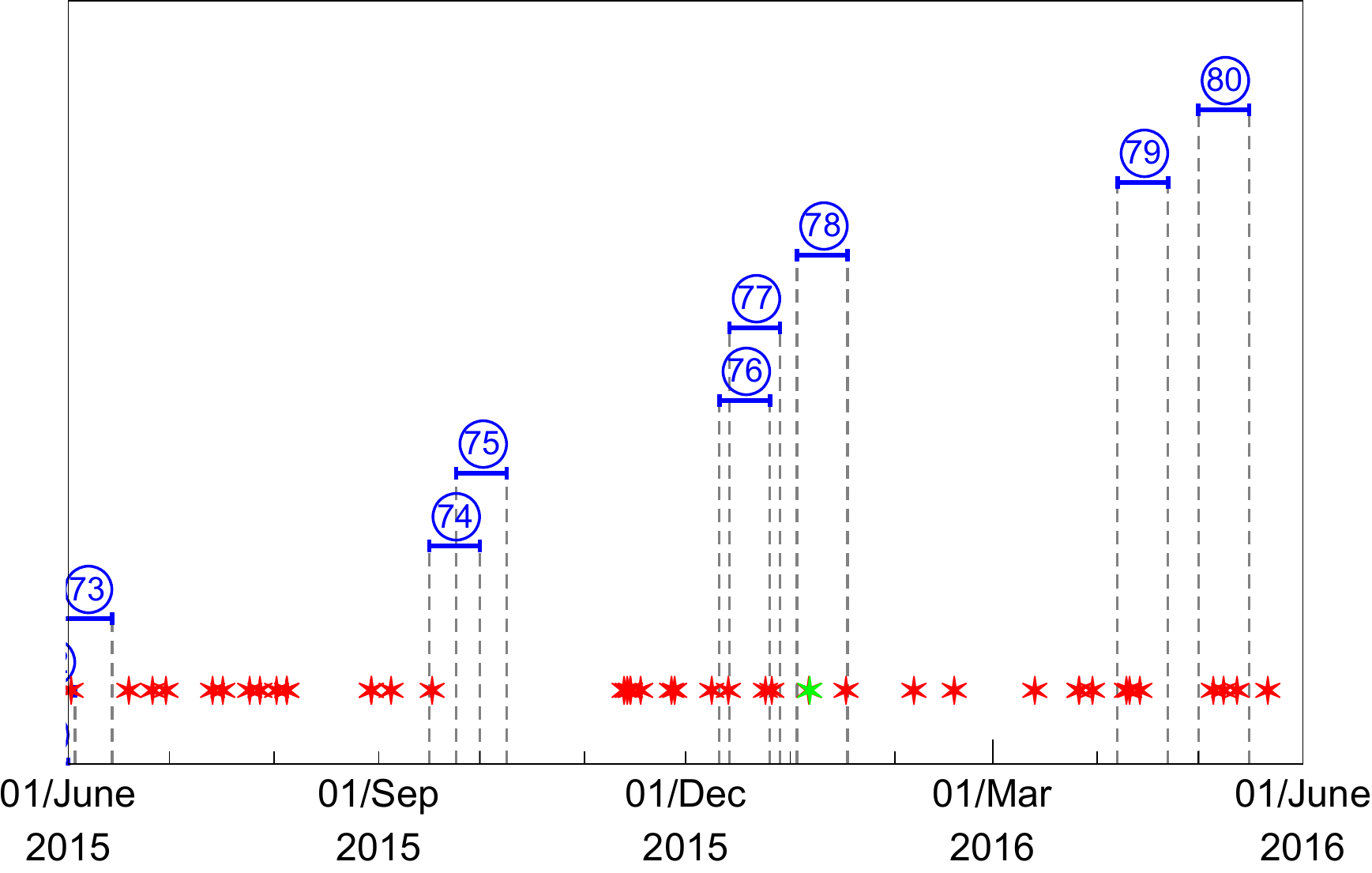}
\label{fig:lum}
}
\caption{\label{fig:dates} The time distribution of the IceCube HESE events and the SNe type Ib/c from June/2010 to June/2016. The blue line segments show the time windows for each neutrino event (the event number is shown above the time window). The black stars show the SNe that happened in the time window of an IceCube event and within its median error circle. The green stars show the SNe happened within the time window of an IceCube event and outside the median error circle, but with a significant directional correlation with the corresponding IceCube events (see section~\ref{sec:method} for more details on the directional correlation). The red stars represent the SNe without any significant directional correlation and/or time correlation.}
\end{figure}
%%%%%%%%%%%%%%%%%%%%%%%%%%%%%%%
%%%%%%%%%%%%%%%%%%%%%%%%%%%%%%%

The 6 years of HESE data consists of $N_\nu=80$ events\footnote{The total number of the events in the 6-yr HESE dataset is 82 where two events, events \#32 and \#55, are coincident events and are not considered in our analysis.} where 22 events are muon-tracks, with median angular resolution $\simeq1^\circ$, and the rest are cascades with larger ($\gtrsim10^\circ$) resolution in the reconstruction of the direction of the incoming neutrino. During the same period $N_{\rm SN}=222$ SNe Ib/c have been registered by several telescopes\footnote{Data are available from Open SNe catalog: \url{https://sne.space}}. There are two date stamps for each SN event: the detection date and the date of its maximum optical brightness. The known correlation between the GRBs and SNe Ib/c suggests the date difference of $\sim13\pm2.3$~days between the GRB's explosion (without emerging jet) and the SN's date of maximum brightness~\cite{Cano:2016ccp}. We will consider this date difference as a Poisson distributed random parameter with mean value $\lambda_T=13$~days. The 90\% confidence interval from Poisson distribution defines a time window for each neutrino event which is given by: $4\leq T_{\rm max,SN} - T_{\nu} \leq 19$, where $ T_{\rm SN,max}$ is the date of SN maximum optical brightness and $T_\nu$ is the date of neutrino event. Notice that we are neglecting the exact detection time of both neutrino and SNe and assume the observation date as a discrete parameter given by the Modified Julian Day of the corresponding event. Figure~\ref{fig:dates} shows the time distribution of the neutrino and SNe events. The blue line segments show the time windows for each neutrino event (the event number is shown above the time window) and the stars show the SNe. The time correlation of the neutrinos and SNe can be seen as the number of stars within the time window of each neutrino event. However, notice that the depicted SNe are \textit{all} the SNe regardless of their relative directions to the neutrino event. The color codes of the stars in Figure~\ref{fig:dates} distinguish between different directional correlation possibilities. The black stars show the SNe which happen in the median error circle of the IceCube event in addition to be in the time window of the event. The green stars show the SNe which happen outside of the error circle of the event, but close enough that still in our analysis a significant correlation will be assigned to the SN and neutrino pair (see section~\ref{sec:method} the Figure~\ref{fig:soverbdir}). The red stars show the rest of the SNe.

\begin{figure}[h!]
\centering
\subfloat{
\includegraphics[width=0.5\textwidth]{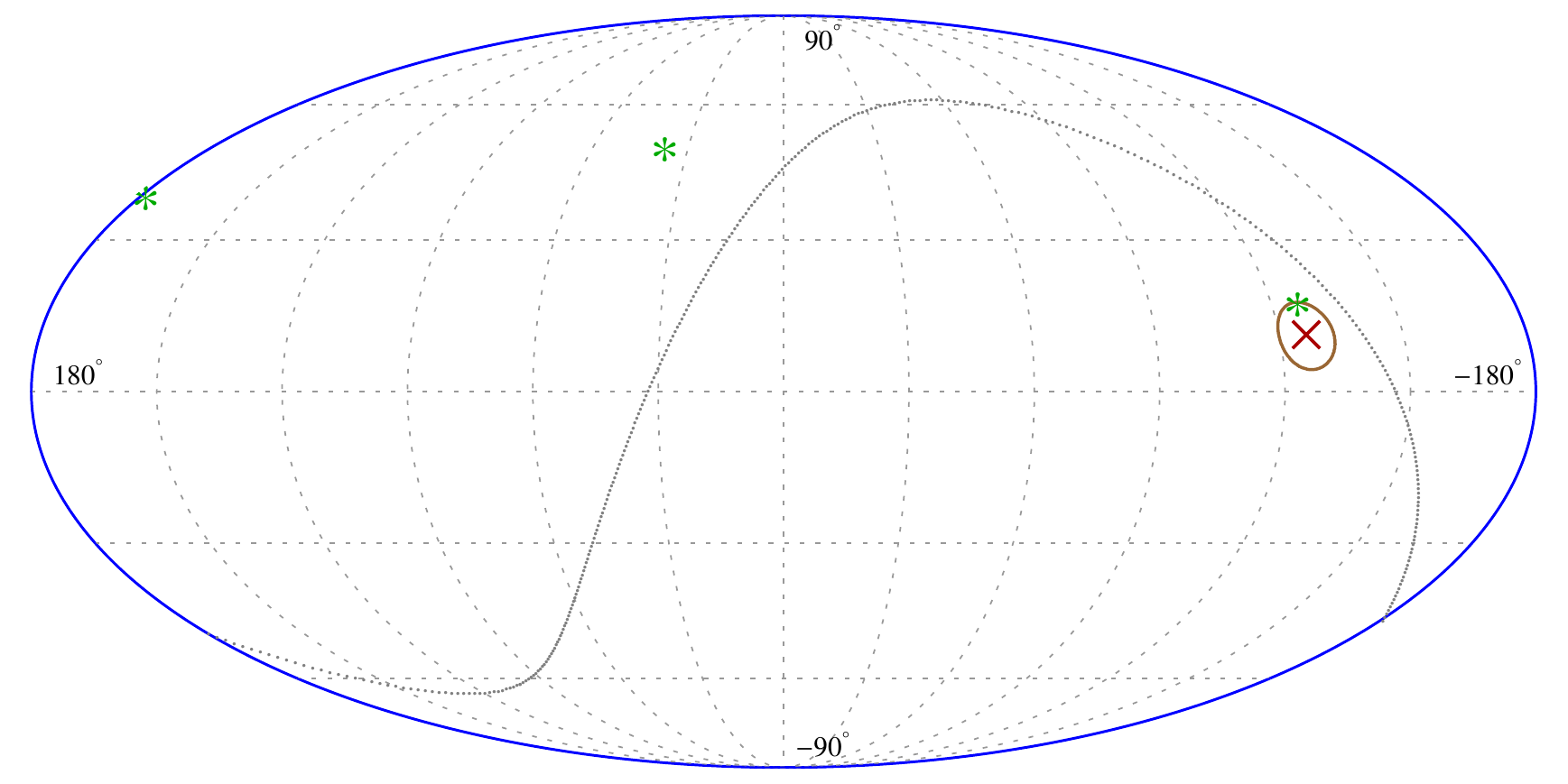}
\label{fig:dndx}
}
\subfloat{
\includegraphics[width=0.5\textwidth]{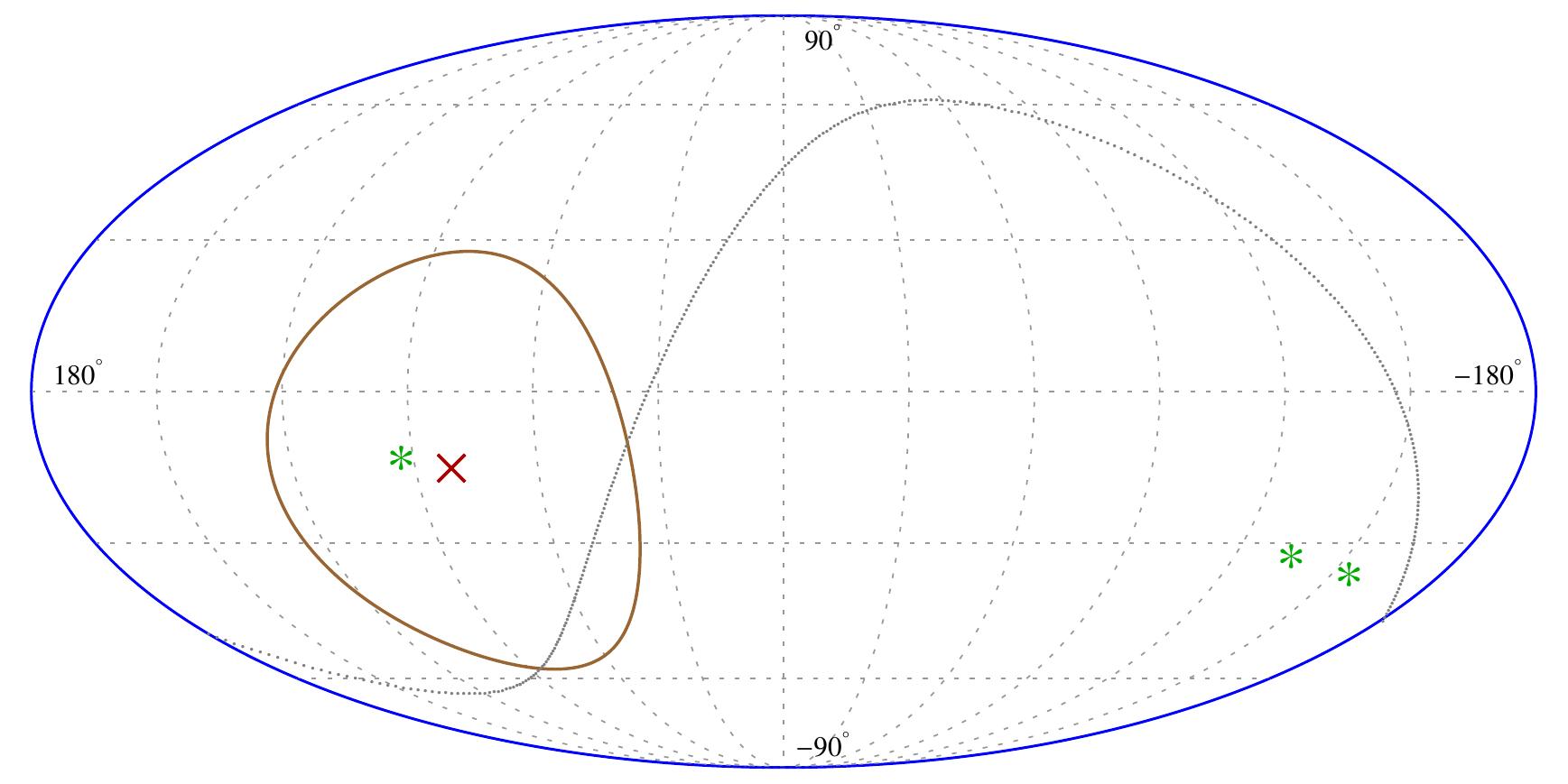}
\label{fig:lum}
}
\caption{\label{fig:eventplot} The skymaps of the IceCube events \#27 (left panel) and \#33 (right panel), in the galactic coordinate. The red $\times$ shows the neutrino event, with the error circle depicted in brown, and the green stars show the SNe occurred within the time window of the neutrino event.}
\end{figure}

As can be seen, out of the 80 neutrino events, 62 neutrino events have at least one SN happening within their corresponding time window. However, out of these 62 events, there are two neutrino events, \#27 and \#33, that have a SN within their median error circles (the two black stars in Figure~\ref{fig:dates}). Figure~\ref{fig:eventplot} shows the skymap of these two events in the galactic coordinate. In this figure the red $\times$ shows the neutrino event, with the error circle depicted in brown, and the green stars show the SNe occurred within the time window of the neutrino event. The neutrino event \#27 is a cascade with the deposited energy $60.2\pm5.6$~TeV and the moderate median angular resolution of $6.6^\circ$, and the accompanied SN is type Ic LSQ12bqn with the luminosity distance $D_L = 430$~Mpc. The neutrino event \#33 also is a cascade with the deposited energy $42.1\pm6.3$~TeV and large median angular resolution $42.7^\circ$, and the correlated SN is the type Ic SN2012gi with $D_L=109$~Mpc. The coincidences with these two SNe are likely to be accidental.  For example, the number of low-luminosity GRBs within an angular patch of $10$~deg and a time window of 10~days is $\sim10-100$, so one neutrino event can be attributed to one of the distant SNe. It is crucial to improve angular resolutions and/or to shorten time windows, and X-ray or MeV gamma-ray observations would be important for the latter purpose. Also, the HESE effective area allows us to estimate the released energy of CRs for theses two events, ${\mathcal E}_{\rm cr}\gtrsim{10}^{53}-{10}^{54}$~erg, which is possible for on-axis jet events but rather extreme.

%%%%%%%%%            figure 2           %%%%%%%%%%%%%
%%%%%%%%%%%%%%%%%%%%%%%%%%%%%%%%%
%%%%%%%%%%%%%%%%%%%%%%%%%%%%%%%
%%%%%%%%%%%%%%%%%%%%%%%%%%%%%%%  

\begin{figure}[h!]
\centering
\includegraphics[width=0.7\textwidth]{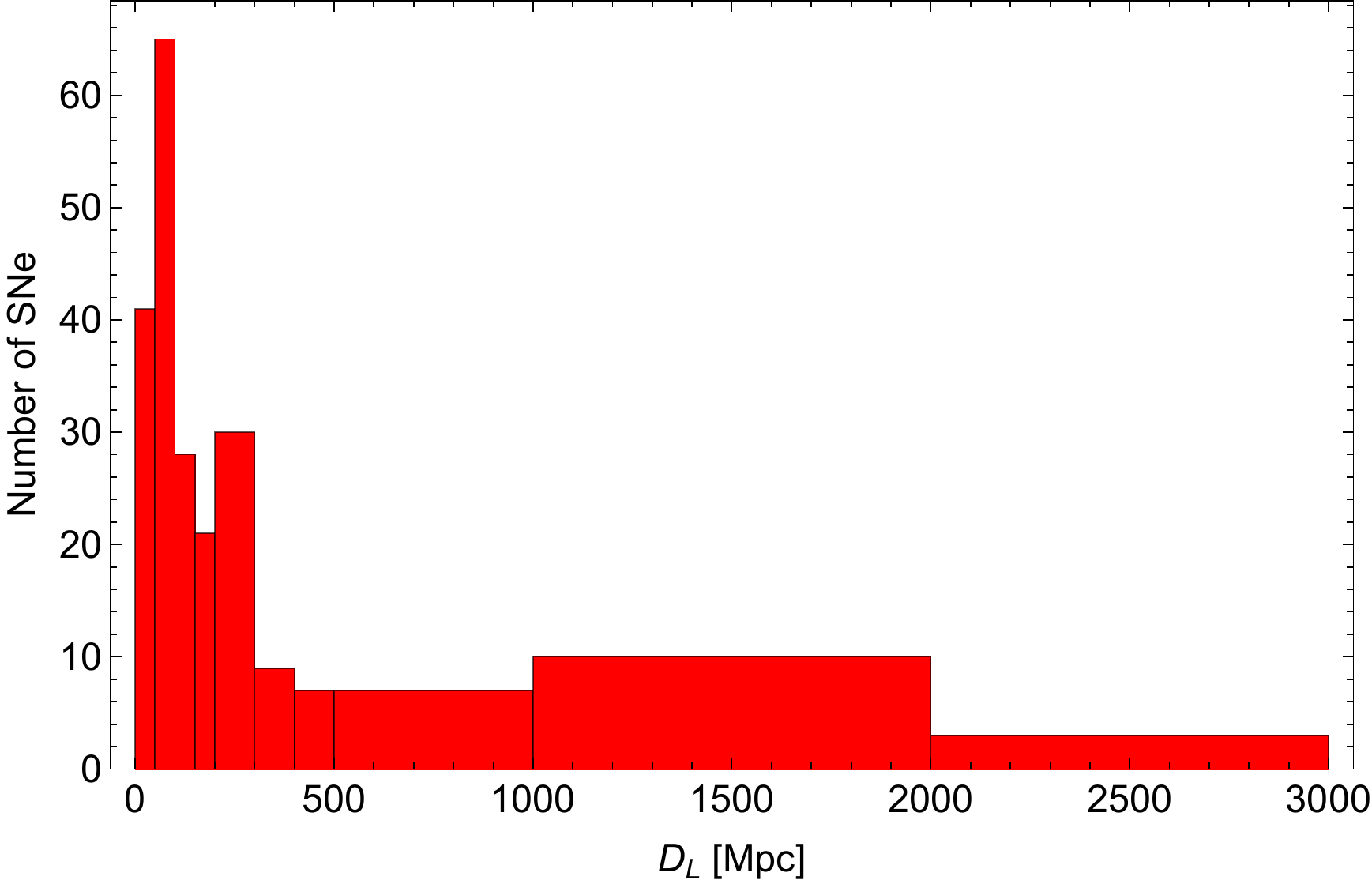}
\label{fig:dndx}
\caption{\label{fig:distance} Histogram of the luminosity distances, $D_L$, of the SNe type Ib/c occurred during June/2010 to June/2016 that will be used in our analysis}
\end{figure}

The luminosity distances, $D_L$, of the observed SNe are of crucial importance since the fluence of each SN scales as $\propto D_L^{-2}$. Using the optical data, the majority of SNe's redshifts are precisely determined. In our sample just for 5 SNe the redshifts (and so the $D_L$) are not measured. For our analysis we will randomly assign luminosity distances to these SNe, chosen from the measured $D_L$s. Figure~\ref{fig:distance} shows the distribution of the luminosity distances of SNe that will be used in our analysis. Notice that the binning of the histogram in Figure~\ref{fig:distance} is not uniform: the bin-width of the first four bins is 50~Mpc, followed by three bins of width 100~Mpc and three bins with the width 500~Mpc.

%%%%%%%%%            figure 3           %%%%%%%%%%%%%
%%%%%%%%%%%%%%%%%%%%%%%%%%%%%%%%%
%%%%%%%%%%%%%%%%%%%%%%%%%%%%%%%
%%%%%%%%%%%%%%%%%%%%%%%%%%%%%%%  

%%%%%%%%%%%%%%%%%%%%%%%%%%
%%%%%%%%%%%%%%%%%%%%%%%%%%
\subsection{Analysis method\label{sec:method}}
%%%%%%%%%%%%%%%%%%%%%%%%%%
%%%%%%%%%%%%%%%%%%%%%%%%%%

In order to quantify the contribution of the choked-jet stripped-envelope SNe to the HESE data set, we perform a stacking analysis on the \textit{neutrinos}. The analysis is very similar to the one performed by the IceCube and Auger collaborations in the search for correlation between the arrival directions of IceCube neutrino events and ultrahigh-energy cosmic rays~\cite{Aartsen:2015dml}. In the following we summarize the details.      

The correlation between HESE and SNe events is quantified by the following likelihood function (in terms of one parameter $n_s$):
\begin{equation}
\log \mathcal{L}(n_s) = \sum_{i=1}^{N_{\rm SN}}\log \left[ \frac{n_s}{N_{\rm SN}} \mathcal{S}^i + \left(1-\frac{n_s}{N_{\rm SN}} \right)\mathcal{B}^i \right]~,
\end{equation}  
where $N_{\rm SN}=222$ is the number of observed SNe Ib/c during the July/2010 to July/2016. The $\mathcal{S}^i$ and $\mathcal{B}^i$ are the signal and background PDFs for the $i^{\rm th}$ SN, respectively, and can be written as:
\begin{equation}
\mathcal{S}^i = \mathcal{S}^i_{\rm dir} \times \mathcal{S}^i_{\rm T}~~~,~~~\mathcal{B}^i = \mathcal{B}^i_{\rm dir} \times \mathcal{B}^i_{\rm T}~,
\end{equation}
where the first and second terms in each equation are, respectively, the directional and temporal contributions to the signal and background PDFs. 

The temporal signal PDF for the $i^{\rm th}$ SN, that is $\mathcal{S}^i_{\rm T}$, can be written as the sum over the mutual signal PDFs of the $j^{\rm th}$ neutrino event and the $i^{\rm th}$ SN: 
\begin{equation}
\mathcal{S}^i_{\rm T} = \sum_{j=1}^{N_\nu} \mathcal{S}^{ij}_{\rm T}~,
\end{equation} 
where $N_\nu=80$. For the $i^{\rm th}$ SN, just the neutrinos falling in the time window $4\leq T_{{\rm max,SN}_i}-T_{\nu_j} \leq 19$ are considered. For these neutrinos the PDF is given by the Poisson probability mass function:
\begin{equation}
\mathcal{S}^{ij}_{\rm T} = e^{-\lambda_T} \frac{\lambda_T^{(T_{{\rm max,SN}_i}-T_{\nu_j})}}{(T_{{\rm max,SN}_i}-T_{\nu_j})!}~,  
\end{equation}
where $\lambda_T=13$~days. 

Similarly, the temporal background PDF, $\mathcal{B}^i_{\rm T}$, can be written as:
\begin{equation}
\mathcal{B}^i_{\rm T} = \sum_{j=1}^{N_\nu} \mathcal{B}^{ij}_{\rm T}~.
\end{equation} 
We assume a uniform distribution within the time window, such that:
\[  \mathcal{B}^{ij}_{\rm T} = \begin{cases} 
      \frac{1}{16} & 4\leq T_{{\rm max,SN}_i}-T_{\nu_j} \leq 19 \\
      0 & {\rm otherwise} 
      \end{cases}~. \]
 
\begin{figure}[htbp]
   \centering
   \subfloat[]{
   \includegraphics[width=0.5\textwidth]{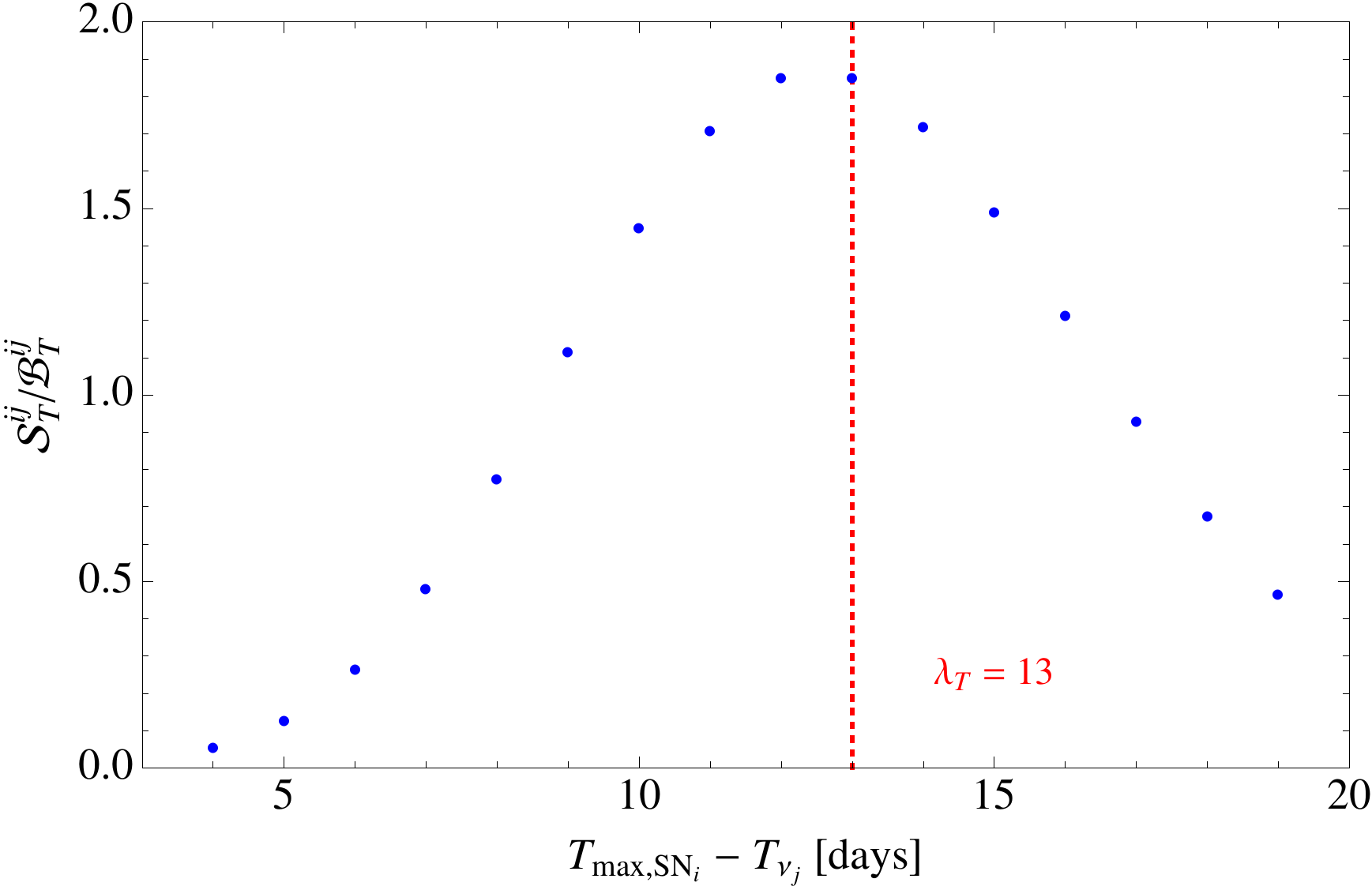}
   \label{fig:soverbtime}
   }
   \subfloat[]{
   \includegraphics[width=0.5\textwidth]{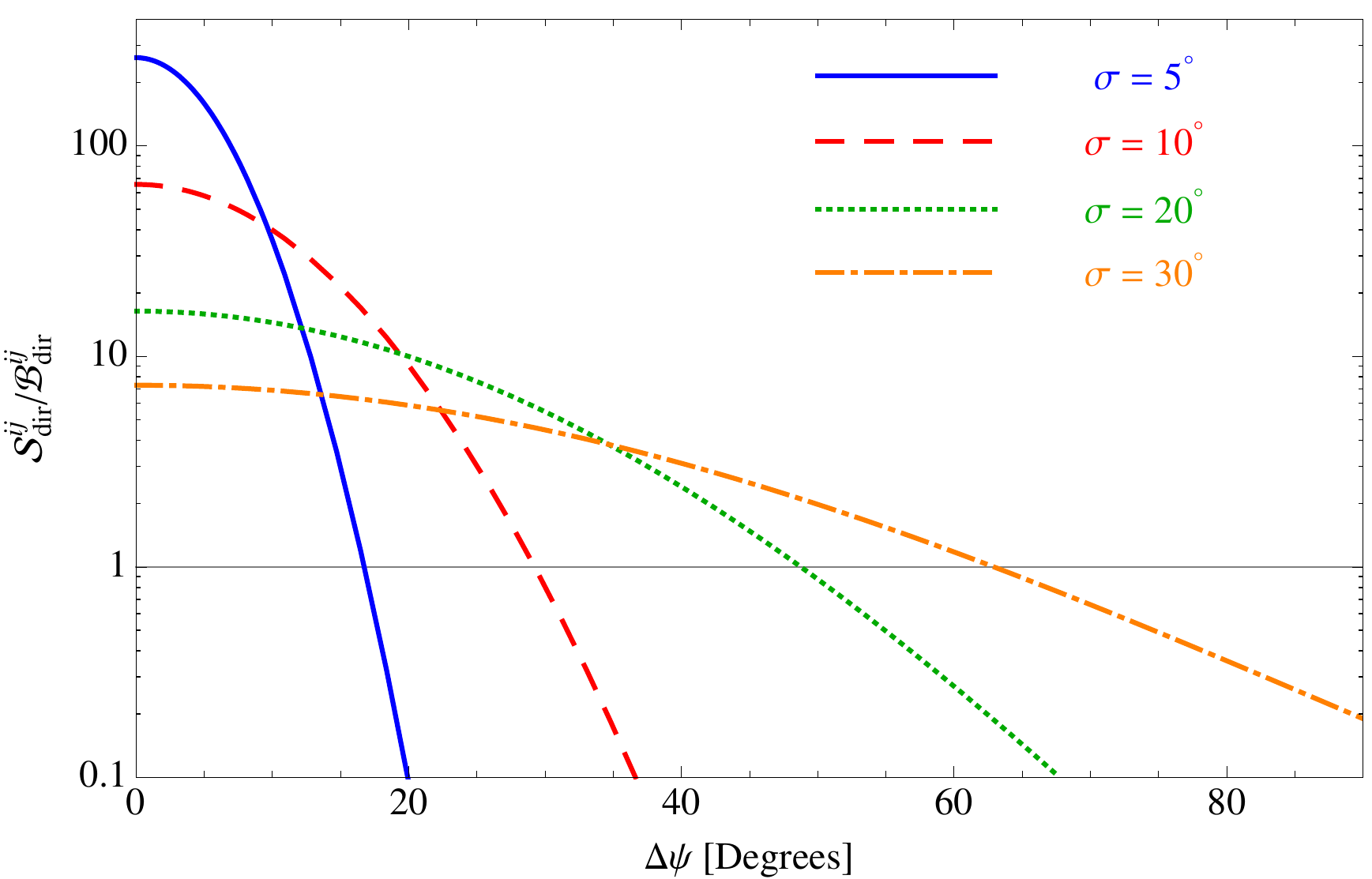}
   \label{fig:soverbdir}
   }
   \caption{The ratio of \textbf{a}) $\mathcal{S}^{ij}_{\rm T} /\mathcal{B}^{ij}_{\rm T}$ as function of the days between $T_{{\rm max,SN}_i}$ and $T_{\nu_j}$; \textbf{b}) $\mathcal{S}^{ij}_{\rm dir} /\mathcal{B}^{ij}_{\rm dir}$ as function of the angular distance $\Delta\psi_{ij}$ for various angular resolutions $\sigma$.}
\end{figure}

Figure~\ref{fig:soverbtime} shows the ratio of $\mathcal{S}^{ij}_{\rm T} /\mathcal{B}^{ij}_{\rm T}$ as function of the day-difference between $T_{{\rm max,SN}_i}$ and $T_{\nu_j}$. As can be seen from the figure, the ratio of $\mathcal{S}^{ij}_{\rm T} /\mathcal{B}^{ij}_{\rm T}$ is larger than one for $9\lesssim T_{{\rm max,SN}_i}-T_{\nu_j}\lesssim16$.

%%%%%%%%%            figure 4           %%%%%%%%%%%%%
%%%%%%%%%%%%%%%%%%%%%%%%%%%%%%%%%
%%%%%%%%%%%%%%%%%%%%%%%%%%%%%%%%%
%%%%%%%%%%%%%%%%%%%%%%%%%%%%%%%%%

The directional signal PDF for the $i^{\rm th}$ SN can be written as
\begin{equation}
\mathcal{S}^i_{\rm dir} = R_i \times\sum_{j=1}^{N_\nu} \mathcal{S}^{ij}_{\rm dir}~,
\end{equation}
where $R_i$ is a factor that takes into account the relative direction-dependence of the detector response for SN detection. Since the SNe we are considering are detected by several experiments we can assume that this factor is equal to one. For the neutrinos that fall within the time window of the $i^{\rm th}$ SN, the $\mathcal{S}^{ij}_{\rm dir}$ is given by (using the Fisher-Bingham or Kent distribution function):
\begin{equation}\label{eq:kent}
\mathcal{S}^{ij}_{\rm dir} =  \frac{\kappa_j}{4\pi\sinh\kappa_j} e^{\kappa_j\mu_{ij}}~,
\end{equation}
where $\kappa_j=1/\sigma_j^2$ and $\mu_{ij}=\cos\Delta\psi_{ij}$. Here $\sigma_j$ is the uncertainty in the direction of the $j^{\rm th}$ neutrino event and $\Delta\psi_{ij}$ is the angular distance between the $j^{\rm th}$ neutrino event and $i^{\rm th}$ SN. 
Note that the IceCube Collaboration recently updated the HESE data which includes some changes in the arrival direction of the neutrinos with respect to the former reported values. However, the new arrival directions are not published yet and just by eye-based comparing one can notice some changes for a few of events. As will be shown, the main power of our analysis comes from the temporal PDFs and so we are safe from these changes.
For the neutrinos falling outside the observation window the signal PDF is zero. The directional background PDF is assumed to be an uniform distribution from all the directions, that is $ \mathcal{B}^{ij}_{\rm dir}=1/4\pi$. Figure~\ref{fig:soverbdir} shows the ratio of $\mathcal{S}^{ij}_{\rm dir} /\mathcal{B}^{ij}_{\rm dir}$ as function of the angular distance $\Delta\psi_{ij}$ for various angular resolutions $\sigma_j$. As can be seen, by increasing the resolution, the width of the curves increase. The Kent distribution in Eq.~(\ref{eq:kent}) assigns significant correlation for $\Delta\psi_{ij}\lesssim3\sigma_j$. For example, for $\sigma_j=10^\circ$, the ratio of $\mathcal{S}^{ij}_{\rm dir} /\mathcal{B}^{ij}_{\rm dir}$ is larger than one for $\Delta\psi_{ij}\lesssim30^\circ$. In the Figure~\ref{fig:dates} the green stars show the SNe with $\mathcal{S}^{ij}_{\rm dir} /\mathcal{B}^{ij}_{\rm dir}$ larger than one.

\begin{figure}[htbp]
   \centering
   \subfloat[]{
   \includegraphics[width=0.51\textwidth]{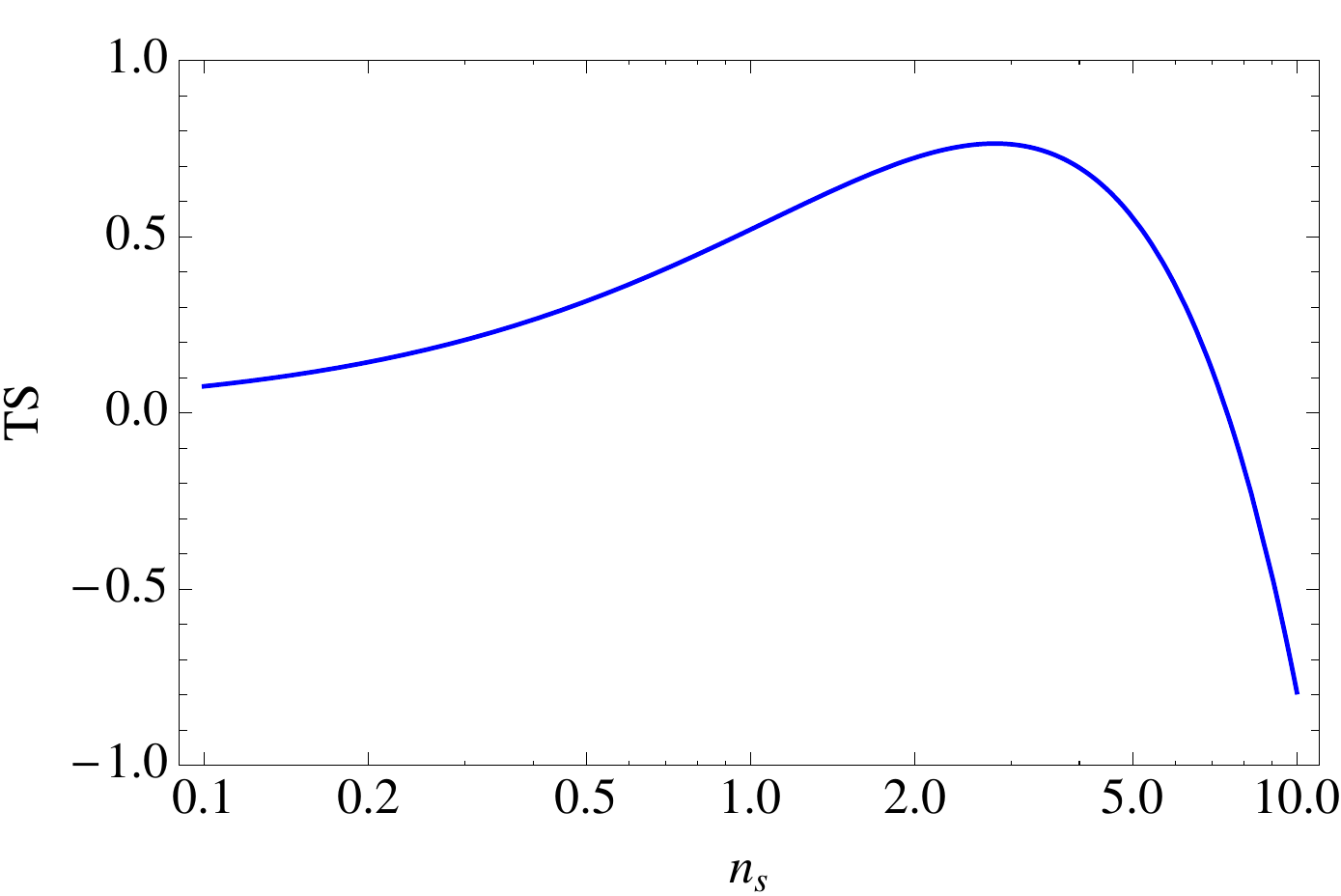}
   \label{fig:ns}
   }
   \subfloat[]{
   \includegraphics[width=0.49\textwidth]{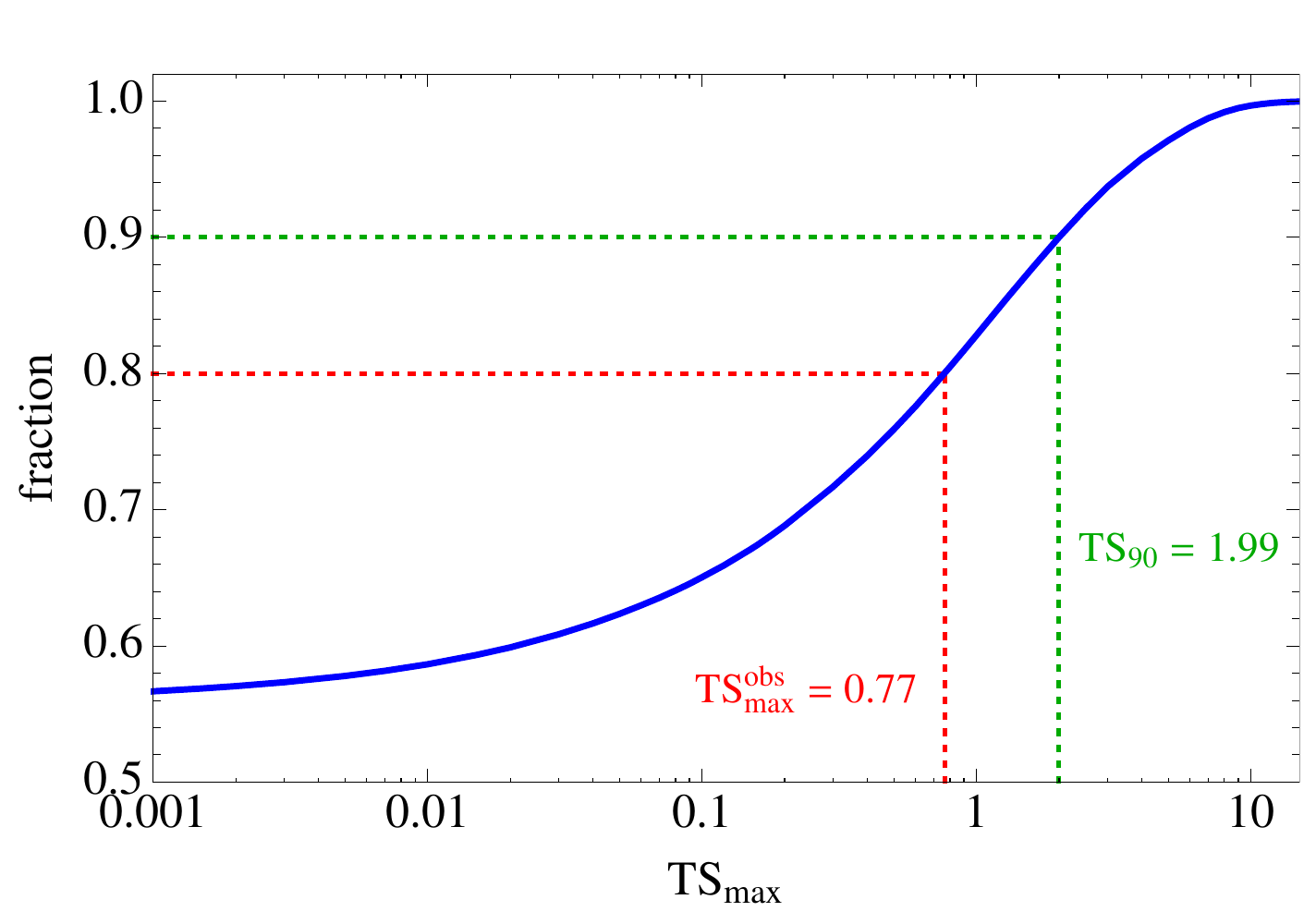}
   \label{fig:ts}
   }
   \caption{\textbf{a}) The value of TS as a function of $n_s$ obtained in our analysis. The maximum value of TS is ${\rm TS}_{\rm max}^{\rm obs} = 0.77$ for $n_s=2.8$; \textbf{b}) The distribution of the maximum TS value for $10^5$ randomly generated sets of SNe. The red vertical dashed line shows the observed maximum TS value, while the green vertical dashed line shows the TS value (that is ${\rm TS}_{90} = 1.99$) which is larger than 90\% of the generated TS values. }
\end{figure}

The test statistics (TS) value is defined by:
\begin{equation}
{\rm TS} = 2\log \left[\frac{\mathcal{L}(n_s)}{\mathcal{L}(n_s=0)}\right]~.
\end{equation}  
The best-fit value of $n_s$ can be obtained by maximizing the TS value. Figure~\ref{fig:ns} shows the TS as a function of the $n_s$ obtained in our analysis. The TS is maximum at $n_s=2.8$ with the value ${\rm TS}_{\rm max}^{\rm obs} = 0.77$. As can be seen, by increasing the $n_s$ the TS decreases rapidly. The $p$-value of the obtained TS value can be estimated by randomly generating SNe events. We generated $10^5$ sets of 222 SNe with random dates and directions and calculated the distribution of the TS values. Figure~\ref{fig:ts} shows the distribution of the maximum TS values obtained in the randomly generated SNe events. As can be seen, the obtained TS$=0.77$ has a $p$-value $\sim20\%$.     

%%%%%%%%%            figure 5           %%%%%%%%%%%%%
%%%%%%%%%%%%%%%%%%%%%%%%%%%%%%%%%
%%%%%%%%%%%%%%%%%%%%%%%%%%%%%%%%%
%%%%%%%%%%%%%%%%%%%%%%%%%%%%%%%%%

%%%%%%%%%%%%%%%%%%%%%%%%%%%%%%%%%
%%%%%%%%%%%%%%%%%%%%%%%%%%%%%%%%%
\section{Upper limits\label{limit}}
%%%%%%%%%%%%%%%%%%%%%%%%%%%%%%%%%
%%%%%%%%%%%%%%%%%%%%%%%%%%%%%%%%%

The large $p$-value obtained in section~\ref{sec:method} points to almost no indication of the correlation between the SNe and HESE events in the current data set. Using this observation, it is possible to place upper limit on the contribution of the SNe to the HESE data. The contribution of the SNe to the observed neutrino events depends on two parameters: \textit{i}) the fraction, $f_{\rm jet}$, of the choked GRBs which have their jets aligned to the Earth; \textit{ii}) the neutrino fluence of a SN explosion at the Earth, or equivalently the energy deposited into cosmic-rays, $\mathcal{E}_{\rm cr}$, in the SN explosion. Note that high-energy neutrino emission can be produced by SN shocks in the dense circumstellar material~\cite{Murase:2010cu,Murase:2017pfe}, in which $f_{\rm jet}\sim1$ is possible. Our analyses are applicable to such cases including the neutrino emission associated with shock breakout~\cite{Kashiyama:2012zn}.
The neutrino fluence $\mathcal{F}_{\nu_\alpha}$ (per flavor) of a SN can be approximated as~\cite{Waxman:1998yy}: 
\begin{equation}
\mathcal{F}_{\nu_\alpha} = \frac{1}{8}\frac{\mathcal{E}_{\rm cr}}{4\pi D_L^2 \mathcal{R}}~,
\end{equation}
where $\mathcal{R}= \ln (E_{\rm cr,max}/E_{\rm cr,min})=18$ (e.g., $E_{\rm cr,max}\sim{10}^{9}$~GeV and $E_{\rm cr,min}\sim10$~GeV) and the factor $1/8$ takes into account the relative energy that the produced neutrinos carry from the parent proton. Assuming $E_{\rm cr}^{-2}$ spectrum for the parent cosmic-rays, the flux of neutrinos per flavor (sum over neutrino and antineutrino) can be written as 
\begin{equation}
\frac{{\rm d}\Phi_{\nu_\alpha}}{{\rm d}E_\nu} = \frac{1}{8}\frac{\mathcal{E}_{{\rm cr}}}{4\pi D_L^2 \mathcal{R}} \frac{1}{\ln(E_{\nu,{\rm max}}/E_{\nu,{\rm min}})}\, E_\nu^{-2}~.
\end{equation}  
Note that the limits on the total CR energy ${\mathcal E}_{\rm cr}$ depends on the spectral index, but limits on the CR energy corresponding to the integration over 10~TeV to 10~PeV neutrino energies are rather insensitive to this assumption~\cite{Senno:2017vtd}.
Taking into account that in the photomeson interaction of the protons $E_\nu=0.05E_{\rm cr}$, we obtain:
\begin{equation}
\frac{{\rm d}\Phi_{\nu_\alpha}}{{\rm d}E_\nu} = \frac{1}{8}\frac{\mathcal{E}_{{\rm cr}}}{4\pi D_L^2 \mathcal{R}^2}\, E_\nu^{-2}~.
\end{equation}
Knowing the flux of neutrinos from a SN at the Earth (assuming the jet is directed to the Earth), we can calculate the expected number of the events in IceCube by
\begin{equation}
\sum_{\alpha} \int_{10 {\rm TeV}}^{10 {\rm PeV}} {\mathcal A}_{\rm eff}^{\nu_\alpha}(E_\nu)\; \frac{{\rm d}\Phi_{\nu_\alpha}}{{\rm d}E_\nu}\, {\rm d}E_\nu~,
\end{equation}
where $\mathcal{A}_{\rm eff}^{\nu_\alpha}$ is the HESE effective area for $\alpha$ flavor~\cite{Aartsen:2013jdh} available in the IceCube webpage~\footnote{\url{https://icecube.wisc.edu/science/data/HE-nu-2010-2012}}, in the energy range of $10$~TeV to $10$~PeV. The expected number of neutrinos from a set of $N_{\rm SN}$ SNe, with the fraction $f_{\rm jet}$ of their jets aligned to the Earth, will be (assuming 1:1:1 neutrino flavor ratio at the Earth)
\begin{equation}
N_{\nu}^{\rm exp} = \sum_{i=1}^{f_{\rm jet}N_{\rm SN}}\sum_{\alpha} \int_{10 {\rm TeV}}^{10 {\rm PeV}} \frac{1}{8}\frac{\mathcal{E}_{\rm cr}}{4\pi D_{L,i}^2 \mathcal{R}^2 E_\nu^2}\, \mathcal{A}_{\rm eff}^{\nu_\alpha}(E_\nu)\, {\rm d}E_\nu~.
\end{equation}  

To place the upper limit on the $\mathcal{E}_{\rm cr}$ and $f_{\rm jet}$ parameters we proceed as following: fixing the directions and dates of the neutrino events to the observed ones in HESE dataset, we generate $N_{\rm SN}=222$ SNe with random directions and dates sampled from uniform distributions over the sky and through the six years data-taking period of HESE, respectively, where each SN has $\mathcal{E}_{\rm cr}$ deposited energy into cosmic-rays. For a fixed value of $f_{\rm jet}$, we force $f_{\rm jet}N_{\rm SN}$ out of the $N_{\rm SN}$ SNe to have directions and dates correlated with $f_{\rm jet}N_{\rm SN}$ randomly chosen neutrino events. The directional correlation between each of the $f_{\rm jet}N_{\rm SN}$ neutrinos and SN events is sampled from the Kent distribution with the angular resolution of the corresponding neutrino event. For the temporal correlation the date of each SN sampled from a Poisson distribution with the mean date difference value of 13 days (ahead) with the corresponding neutrino event. This process realized $10^5$ times for each set of the fixed values of $\mathcal{E}_{\rm cr}$ and $f_{\rm jet}$. For each realization we calculate the maximum TS value using the likelihood method described in section~\ref{sec:method}. The rejection confidence level of a set of $\mathcal{E}_{\rm cr}$ and $f_{\rm jet}$ values is defined as the percentage of the realizations that have a maximum TS value larger than 90\% of the generated TS values in the background only hypothesis; {\it i.e.}, when no correlation is introduced between neutrino and SN events. This value, denoted by ${\rm TS}_{90}$ and depicted by the green vertical dashed line in figure~\ref{fig:ts}, is ${\rm TS}_{90}=1.99$. 

\begin{figure}[ht!]
   \centering
   \includegraphics[width=0.75\textwidth]{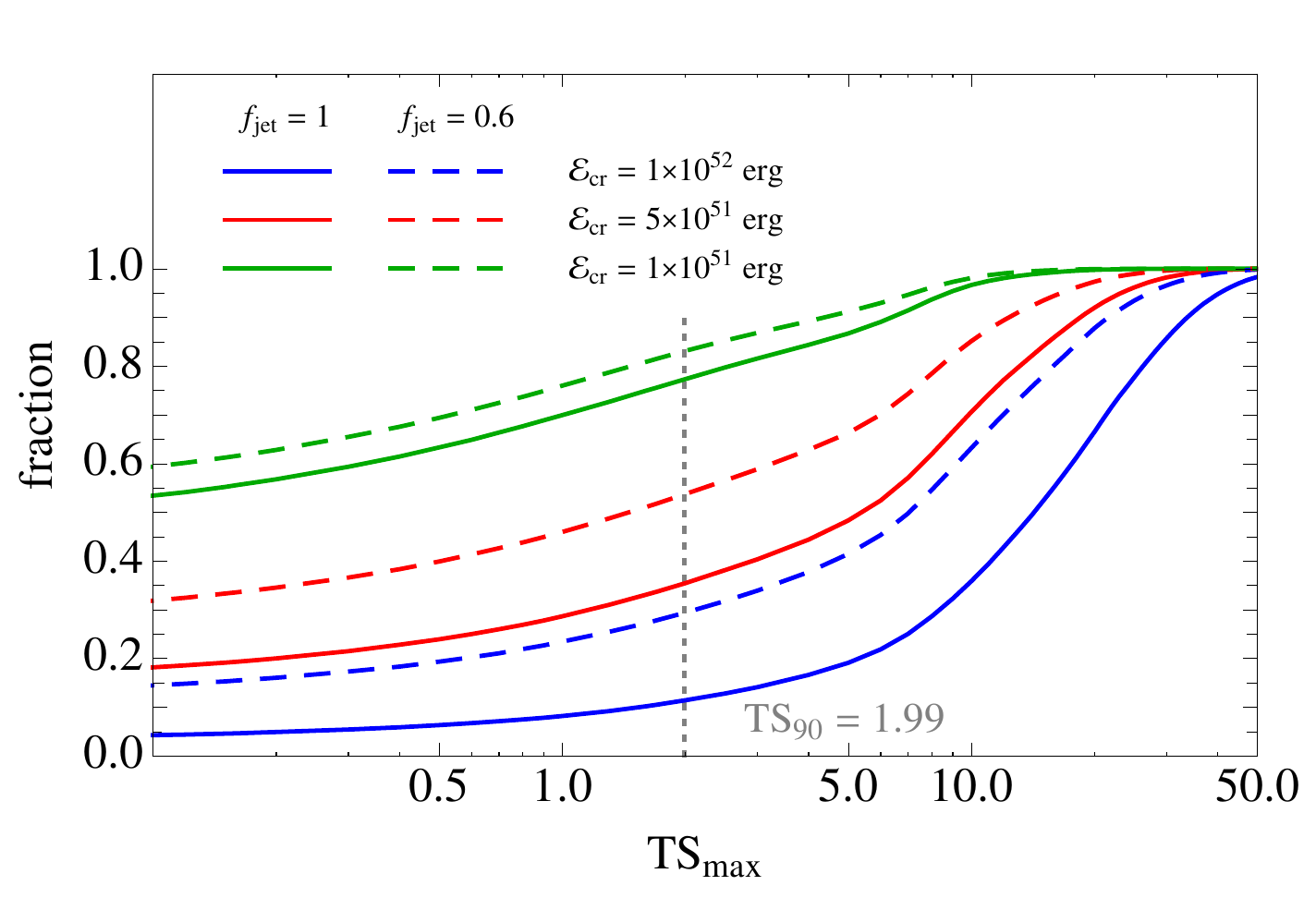}
   \caption{\label{fig:cdfs}The cumulative distribution of ${\rm TS}_{\rm max}$ values for $10^{5}$ realizations randomly generated SNe sets assuming $f_{\rm jet}=1$ (0.6), shown by solid (dashed) curves, and $\mathcal{E}_{\rm cr}=10^{52}$, $5\times10^{51}$ and $10^{51}$ erg, depicted respectively by blue, red and green colors. The gray dotted vertical line shows the ${\rm TS}_{90}=1.99$.
   }
\end{figure}

\begin{figure}[ht!]
   \centering
    \includegraphics[width=0.7\textwidth]{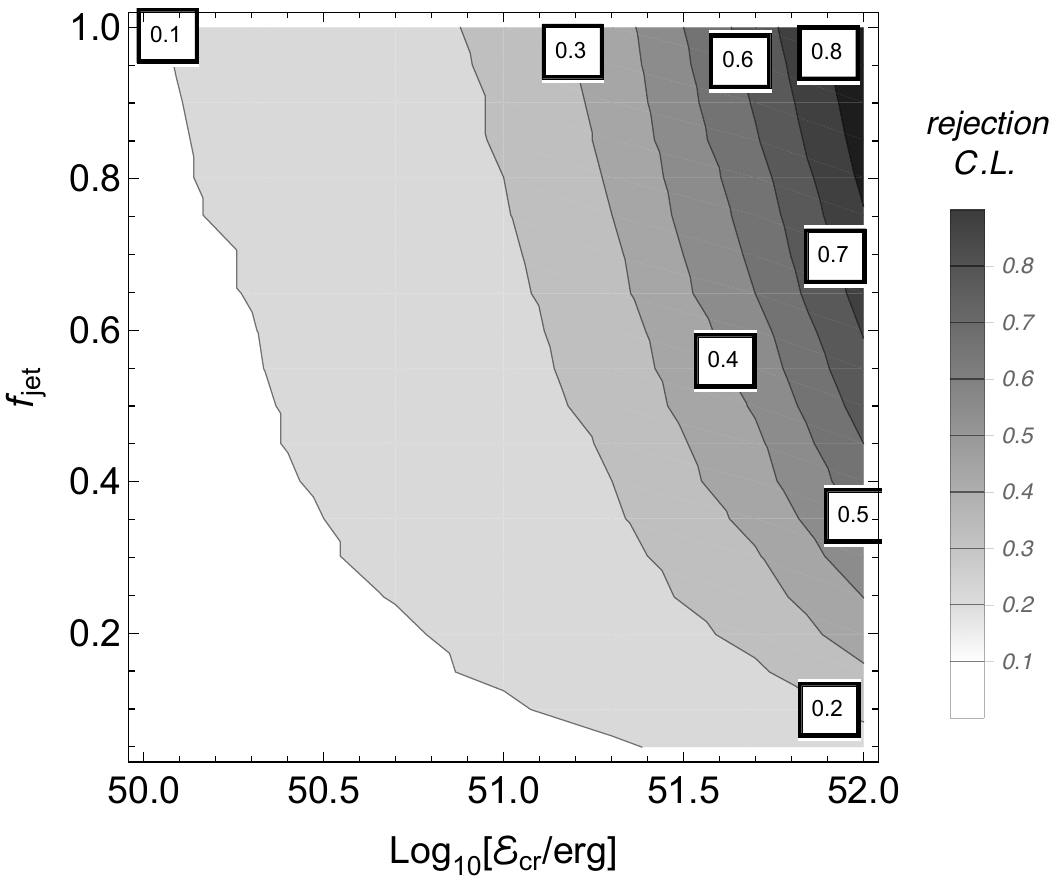}
   \caption{\label{fig:heat}The heat plot in the parameter space of $f_{\rm jet}$ and $\mathcal{E}_{\rm cr}$. The gray scale shows the rejection C.L. (the grayer color corresponds to higher rejection C.L.).
   The right top corner corresponds to the 88\% C.L. limit.
   }
\end{figure}

Figure~\ref{fig:cdfs} shows the cumulative distribution of ${\rm TS}_{\rm max}$ values obtained in the $10^5$ realizations of the process described above, depicted by solid (dashed) curves for $f_{\rm jet}=1$ (0.6) with colors blue, red and green respectively for $\mathcal{E}_{\rm cr}=10^{52}$, $5\times10^{51}$ and $10^{51}$ erg. The gray dotted vertical line shows the ${\rm TS}_{90}=1.99$. The rejection C.L. of a given set of $f_{\rm jet}$ and $\mathcal{E}_{\rm cr}$ is given by the percentage of the ${\rm TS}_{\rm max}$ values that are larger than ${\rm TS}_{90}$; which can be read from figure~\ref{fig:cdfs} as the corresponding fraction value of the intersection between the gray dotted line and cumulative distribution curve. Scanning over the parameter space of $f_{\rm jet}$ and $\mathcal{E}_{\rm cr}$, figure~\ref{fig:heat} shows the heat plot which gives the rejection C.L. of each point. For example, in figure~\ref{fig:heat}, the top right corner (corresponding to $f_{\rm jet}=1$ and $\mathcal{E}_{\rm cr}=10^{52}$~erg) is rejected at 88\% C.L. 

%%%%%%%%%            figure 6           %%%%%%%%%%%%%
%%%%%%%%%%%%%%%%%%%%%%%%%%%%%%%%%
%%%%%%%%%%%%%%%%%%%%%%%%%%%%%%%%%
%%%%%%%%%%%%%%%%%%%%%%%%%%%%%%%%%

%%%%%%%%%            figure 7           %%%%%%%%%%%%%
%%%%%%%%%%%%%%%%%%%%%%%%%%%%%%%%%
%%%%%%%%%%%%%%%%%%%%%%%%%%%%%%%%%
%%%%%%%%%%%%%%%%%%%%%%%%%%%%%%%%%

\section{Discussion}\label{discussion}
Ref.~\cite{Senno:2017vtd} constrained high-energy neutrino emission from choked-jet SNe, using the one-year sample of muon neutrinos in IceCube. For $f_{\rm jet}\rightarrow1$, the 90\% C.L. upper limit on the total CR energy is ${\mathcal E}_{\rm cr}\lesssim{10}^{52}$~erg, which is comparable to the limit obtained in this work. Because the time coincidence drastically reduces the atmospheric backgrounds, poorer angular resolutions of HESE events do not cause a serious problem for our purpose.  
The fluence of stacked SNe is naively written as ${\mathcal F}_{\nu} N_{\rm SN}^{\rm corr}$, where $N_{\rm SN}^{\rm corr}$ is the number of SNe correlated with the neutrino events. 
Then, for given energy $E_\nu$ and observation time ${\mathcal T}_{\rm obs}$, the upper limit roughly scales as ${\mathcal E}_{\rm cr}^{\rm lim}\propto {\mathcal A}_{\rm eff}^{-1}{\mathcal T}_{\rm obs}^{-1}$. The HESE effective area~\cite{Aartsen:2013jdh} is worse than the upgoing muon neutrino one~\cite{Aartsen:2016oji}, which is compensated by the larger data set used in our analysis (${\mathcal T}_{\rm obs}=6~{\rm yr}$). 

Stripped-envelope SNe harboring choked jets have been discussed as possible candidates for the dominant origin of IceCube neutrinos even in the 10-100~TeV range~\cite{Murase:2013ffa}.  
If these SNe are responsible for 100\% of the diffuse neutrino flux, the required CR luminosity density is $Q_{\rm cr}\sim{\rm a~few}\times{10}^{45}~{\rm erg}~{\rm Mpc}^{-3}~{\rm yr}^{-1}$ for the $E_{\rm cr}^{-2}$ spectrum~\cite{Murase:2015xka}.
Then the upper limit of ${\mathcal E}_{\rm cr}\lesssim{10}^{52}~f_{\rm jet}^{-1}~{\rm erg}$ implies that the rate density is $\rho\gtrsim{\rm a~few}\times{10}~(f_{\rm jet}/0.1)~{\rm Gpc}^{-3}~{\rm yr}^{-1}$. 
Note that the local rate densities of stripped-envelope SNe and broadline SNe Ib/c are $R\sim20000~{\rm Gpc}^{-3}~{\rm yr}^{-1}$ and $R\sim2000~{\rm Gpc}^{-3}~{\rm yr}^{-1}$, respectively.  
If only a fraction ($f_{\rm jet}\approx\theta_{\rm jet}^2/2$) of the SNe have jets beamed toward us, the corresponding apparent rate densities are $\rho=f_{\rm jet}R\sim900~{(\theta_{\rm jet}/0.3)}^2~{\rm Gpc}^{-3}~{\rm yr}^{-1}$ and $\rho\sim90~{(\theta_{\rm jet}/0.3)}^2~{\rm Gpc}^{-3}~{\rm yr}^{-1}$, respectively. Here $\theta_{\rm jet}$ is the jet opening angle. Thus these choked-jet SN models can provide viable explanations for the diffuse neutrino flux without violating the stacking limits. 

Note that the diffuse neutrino flux is determined by the product of the isotropic-equivalent CR energy ${\mathcal E}_{\rm cr}$ and the apparent rate of choked-jet SNe $\rho$. Thus, for a given rate of low-power GRBs, one can constrain the multiplication factor by choked jets, $f_{\rm cho}$~\cite{Murase:2013ffa,Senno:2015tsn}. In principle, the degeneracy that exists in the diffuse flux constraints can be broken by exploiting the spectral information~\cite{Denton:2017jwk}. However, this is model dependent, because the spectrum is affected by the unknown distribution of the bulk Lorentz factor and mechanisms for the maximum neutrino energy. Note that the high-energy neutrino cutoff can be caused by not only meson or muon cooling processes but also neutrino absorption in the progenitor~\cite{Murase:2013ffa}, and the spectral shape could be affected by possible reacceleration of mesons and muons~\cite{Murase:2011cx}. The baryon loading could also depend on the Lorentz factor because baryon-rich jets are likely to be slower. Also, in the scenario where jets are choked in the circumstellar material~\cite{Senno:2015tsn} rather than inside the progenitor star, one would need to take into account the distribution of the circumstellar material mass and radius. 

Another piece of important information comes from multiplet and auto-correlation analyses~\cite{Kowalski:2014zda,Ahlers:2014ioa,Murase:2016gly,Aartsen:2014ivk,Aartsen:2017kru}. 
The number of multiplet sources gives a lower limit on the effective rate density of the dominant neutrino sources as~\cite{Senno:2016bso}
\begin{equation}
\rho\gtrsim60~{\rm Gpc}^{-3}~{\rm yr}^{-1}~\frac{q_L^2{(\Delta\Omega/2\pi)}^2{({\mathcal T}_{\rm IC}/6~{\rm yr})}^2}{{(\xi_z/3)}^3{\mathcal F}_{\rm lim,-3.9}^3},
\end{equation}
where $q_L$ is the luminosity-dependent correction factor, $\Delta\Omega$ is the observed solid angle, and $\xi_z$ represents the redshift evolution of the sources~\cite{Murase:2016gly}. 
The above limit is also consistent with the latest limits by Ref.~\cite{Aartsen:2018fpd}. 
At present, neither stacking nor multiplet search gives a sufficiently strong constraint on the choked-jet SN models as the origin of IceCube neutrinos.

\section{Summary}\label{summary}
We searched for temporal and spatial coincidences between high-energy neutrino events in the six-year HESE data and SNe Ib/c that occurred in this time period. 
We did not find any significant correlation, by which we placed upper limits on the total neutrino energy ${\mathcal E}_{\rm cr}$. 
The 90\% C.L. limit for $f_{\rm jet}\rightarrow1$, ${\mathcal E}_{\rm cr}\lesssim10^{52}$~erg, is comparable to that obtained by the independent stacking analysis based on the one-year muon neutrino data.  
Our result demonstrates that meaningful constraints for transients with sufficiently short durations can also be obtained by using shower events with poorer angular resolutions.

The current upper limits are not yet sufficient to exclude the relevant parameter space of choked-jet SN models. However, the situation will be drastically improved in near future. 
First, the number of neutrino events does not have to be restricted to HESE events. One can use the shower data with lower energies for this kind of stacking search. Furthermore, IceCube-Gen2~\cite{Aartsen:2014njl} is expected to have a volume about ten times larger than that of the current IceCube, by which the number of signals can also be about ten times larger with a similar observation time. KM3Net~\cite{Adrian-Martinez:2012qpa} with a better angular resolution for shower events would also be useful for this kind of study. We have checked that improving the angular resolution of shower events to $\sim5^\circ$ can remedy the obtained rejection C.L. by $\sim10\%$. Second, the current catalog of SN Ib/c are highly incomplete, so that $N_{\rm SN}^{\rm corr}$ is quite limited. The number of SN samples will be increased with future SN surveys via, e.g., Zwicky Transient Factory, Kiso Tomo-e Gozen, and Large Synoptic Survey Telescope. Critical constraints on the choked-jet SN models will be obtained if we can achieve ${\mathcal E}_{\rm cr}\lesssim10^{50}-{10}^{51}$~erg in the limit of $f_{\rm jet}\rightarrow1$.

In addition to such stacking analyses, as proposed by Ref.~\cite{Murase:2006mm}, ``follow-up'' searches for electromagnetic counterparts of neutrino events can provide a powerful way to discover the sources of high-energy neutrinos, and choked-jet SNe could be found by optical follow-up observations. The feasibility of such high-energy neutrino triggered campaigns has recently been demonstrated by the measurements of the flaring blazar TXS 0506+056 followed by the discovery of the high-energy neutrino event, IceCube-170922A~\cite{Aartsen2018blazar1,Keivani:2018rnh}. 
Along this line, the Astrophysical Multimessenger Observatory Network~\cite{Smith:2012eu} can play a role in the real-time detections of neutrino transients especially when the sources are accompanied by X-ray or gamma-ray emission. Indeed, jet-driven SNe may possess trans-relativistic ejecta, which are expected to naturally cause short-duration X-ray and gamma-ray emission at the shock breakout. 

%%%%%%%%%%%%%%%%%%%%%%%%%%%%%%%%%%%%%%%%%%%%%%%%%%
%%%%%%%%%%%%%%%%%%%%%%%%%%%%%%%%%%%%%%%%%%%%%%%%%%

%\medskip
\acknowledgments
We thank John Beacom, Mauricio Bustamante, and Peter Denton for useful comments. 
A.~E. thanks the partial support by the CNPq grant No.~310052/2016-5 and resources from FAPESP Multi-user Project 09/54213-0. The work of K.~M. is supported by NSF Grant No. PHY-1620777 and the Alfred P.~Sloan Foundation. This research was supported by the Munich Institute for Astro- and Particle Physics (MIAPP) of the DFG cluster of excellence "Origin and Structure of the Universe".

%%%%%%%%%%%%%%%%%%%%%%%%%%%%%%%%%%%%%%%%%%%%%%%%%%
%%%%%%%%%%%%%%%%%%%%%%%%%%%%%%%%%%%%%%%%%%%%%%%%%%
\bibliographystyle{revtex}
\bibliography{kmurase}

\end{document}